\documentclass[11pt,preprint]{aastex}  
 
\def\etal{{et al.~}}
\def\eg{e.g.,~}
\def\ie{i.e.,~}

\def\kms{{\rm\,km\,s^{-1}}}

\def\Reff{\rm R_{eff}}

\def\vlos{v_{\rm los}}

\def\deg{^{\rm o}}
 
\def\spose#1{\hbox to 0pt{#1\hss}}
\def\lta{\mathrel{\spose{\lower 3pt\hbox{$\sim$}}
    \raise 2.0pt\hbox{$<$}}}
\def\gta{\mathrel{\spose{\lower 3pt\hbox{$\sim$}}
    \raise 2.0pt\hbox{$>$}}}

\begin{document}

\slugcomment{{\it The Astrophysical Journal; resubmitted January 17, 2001}}
  
\title{The kinematics of 3:1--merger remnants and the formation of low--luminosity
elliptical galaxies}

\author{N.~Cretton, T.~Naab, Hans--Walter Rix, A.~Burkert}
\affil{Max-Planck-Institut f\"ur Astronomie, K\"onigstuhl 17, Heidelberg, Germany}

\clearpage      
                                           
\baselineskip=14pt 
%\baselineskip=30pt 

%------------------------------------------------------------------------
% Abstract
%

\begin{abstract}   
 
 We test the formation of low--luminosity elliptical galaxies through
 collisionless mergers of unequal--mass disk galaxies. The kinematic
 properties of a small survey of simulated merger end--products with
 initial disk mass ratios of 3:1 is compared to a sample of seven
 low--luminosity galaxies observed by Rix \etal that were chosen
 photometrically to be "ellipticals". In this paper, we go beyond a
 comparison in terms of global properties (using one number to
 characterize a model or a galaxy \eg $\langle a_4 \rangle$,
 ellipticity at some fixed radius or central velocity dispersion) and
 examine the detailed kinematics as function of galactocentric
 distance.  The merger remnants are ``observed'' through a slit along
 the major and minor axis, using a pixel binning and slit width
 similar to the one used during the spectroscopic observations.
 Inside each bin, we determine the line--of--sight velocity
 distributions and parametrize them with Gauss--Hermite functions. We
 compare the rotational support of the merger remnants, \ie the ratio
 of the mean line--of--sight velocity $v$ to the velocity dispersion
 $\sigma$ along the major axis, the normalized rotation on the minor
 axis and the major axis Gauss--Hermite moments $h_3$, to that of the
 observed galaxies.

 The N--body remnants are very flattened when viewed edge--on
 ($\langle \epsilon \rangle \sim 0.6$) and should be inclined before
 making a fair comparison with the Rix \etal data set (which has
 $\langle \epsilon \rangle \sim 0.3$). When the merger remnants are
 appropriately inclined, their $v/\sigma$ profiles rise slower than
 the observed ones: $v/\sigma_{\rm merger}$ is in the range [0.1, 0.8]
 at 1 $\Reff$ and [0.2, 1.6] at 2 $\Reff$, whereas the $v/\sigma_{\rm
 observed}$ span the intervals [0.8, 2.0] at 1 $\Reff$ and [1.4, 3.5]
 at 2 $\Reff$. Note that even when the remnants are viewed edge--on,
 the $v/\sigma$ profiles do not match the observations. 

The detailed comparison of the observations with our set of purely
 collisionless 3:1--merger remnants shows that these objects and
 low--luminosity ellipticals do not have similar kinematic profiles.
 This suggests that this kind of dissipationless merger (or mergers
 with more even masses, \eg mass ratios of 2:1 or 1:1) is not likely to 
 be the dominant formation channel for low--luminosity elliptical galaxies.

\end{abstract}


\clearpage    

%------------------------------------------------------------------------
% Beginning of main text
%

\section{Introduction}
\label{s:intro}
 
Elliptical galaxies can be divided into two classes according to their
morphology and kinematic properties.  The giant ellipticals are
generally boxy (probably triaxial) with shapes supported by
anisotropy. They have slow rotation on the major axis with often
comparable minor axis rotation, and occasionally kinematically
decoupled cores.  The intermediate and low-luminosity ellipticals are
usually disky, isotropic rotators. They are mostly supported by
rotation and have little minor axis velocities (Bender 1988; Bender,
D\"obereiner \& M\"ollenhoff 1988; Kormendy \& Bender 1996; Rix \etal
1999).  These two families also show correlations with the inner slope
of the surface brightness profile: galaxies with steep inner cusps are
on average small disky objects and galaxies with shallow or constant
cores are the giant boxy ellipticals (Jaffe \etal 1994, Faber \etal
1997, Carollo \etal 1997).

 Barnes (1998) proposed that fast rotating elliptical galaxies could originate
 from collisionless merger of unequal--mass disks. On the other hand,
 equal--mass mergers lead to slowly rotating, pressure supported
 objects resembling giant elliptical galaxies (Barnes 1992).
 Recently Naab, Burkert \& Hernquist (1999, hereafter NBH) took
 Barnes' idea one step further and showed that equal and unequal--mass
 mergers could reproduce most of the observed correlations mentioned
 above if one would analyse projections of the merger remnants using
 many random viewing angles (see their figure 3). They showed that the
 boxy family of galaxies could be explained by equal--mass mergers of
 disk galaxies (with bulges and dark halos), whereas the disky family
 was originating from 3:1 mergers, \ie in which the big disk galaxy is
 3 times more massive than the small one.

 However, NBH's conclusion is based only on global properties (average
 $a_4$, ellipticity measured at one radius, or central velocity
 dispersion) and does not compare kinematic quantities {\em as
 function of radius}. In this paper, we perform such a comparison,
 using a setup similar to the one used during the observations of
 Rix \etal 1999 (hereafter R99), where kinematics along the principal 
axes were obtained, extending well beyond $\Reff$. R99 selected these 
elliptical galaxies only on the basis of their low--luminosities 
(${\rm M_B} \ge -19.5$), \ie no prior kinematic information or 
degree of diskiness was used. 

In section \ref{s:models}, we summarize our merger models and their
initial conditions. In section \ref{s:viewing_merger}, we describe how
we extract the kinematic quantities of the merger remnants. We discuss
the results of a small survey of 16 different merger remnants in
section \ref{s:discussion}.  We perform a comparison between models
and observations using the rotational support (\ie $v/\sigma$ along
the major axis, the amount of minor to major axis rotation and the
Gauss-Hermite moment $h_3$ along the major axis. All these quantities
are compared as function of distance from the center. The edge--on and
inclined merger models fail to match the observed quantities of the
R99 sample.

 In a similar study, Bendo \& Barnes (2000, hereafter BB) examined 8
 equal--mass merger remnants and 8 unequal--mass (3:1) merger
 remnants. Their unequal--mass merger models display similar major
 axis kinematic profiles than ours. However, the authors reach
 different conclusions from the comparison with the R99 data set,
 namely that "unequal--mass merger mergers can produce the same
 relations between $v/\sigma$ and radius".  In this paper, we do a
 more direct comparison of N--body merger remnants with this data
 set. We compare not only edge--on but also inclined remnants that
 match the apparent ellipticities of the observed galaxies. We
 conclude that, on average, the kinematics of the simulated 3:1
 mergers do not reproduce those of low--luminosity galaxies. Therefore
 we suggest that the 3:1 collisionless mergers are probably not a
 major mechanism responsible for the bulk formation of these objects.

\section{The merger models}
\label{s:models}

 We briefly summarize some properties of the initial disk model (see
 Hernquist 1993 and NBH for more details). Each disk galaxy
consists of an exponential disk, a Hernquist (1990) spherical bulge
and a pseudo--isothermal spherical dark halo with 5.8 times the mass
of the disk. In our "low--resolution" simulations, the more massive
galaxy has 6666 bulge particles, 20000 disk particles and 40000 dark
halo particles, while the less massive galaxy has a third of the number of
particles in each component. In two "high--resolution" cases, we rerun
models with three times as many particles. The N--body computations were
performed using a direct summation code on the special purpose hardware
GRAPE--3 AF (GRavity PipE, Sugimoto \etal 1990).

Before the encounter, the two disk galaxies follow a quasi--parabolic
orbit with an initial separation of 30 scale--lengths of the massive
exponential disk. The pericentric
approach is taken as 2 disk scale--lengths. Once the orbits and the
masses of the two initial spiral galaxies are fixed, we still have 2
free parameters for each galaxy: the inclination between the galaxy
orbital plane and its spin plane (angle $i$) and the argument of
pericenter (angle $\omega$), \ie the angle between the line of nodes
and the pericentric distance (see Figure 6a of Toomre \& Toomre, 1972).
Table~\ref{t:initial_angles} summarizes our choice for these angles:
$i_1$, $\omega_1$ for the more massive spiral galaxy and $i_2$,
$\omega_2$ for the less massive one.

\section{Observing the merger remnant}
\label{s:viewing_merger}

 We analyse the kinematic structure of the merger remnants 10
dynamical times after the merger has been completed, so that they had
enough time to settle into an equilibrium state.  We plot the
kinematic quantities of the remnant in units of the half--light radius
$\Reff$, \ie the projected radius on the plane of the sky of the
circle containing half of the luminous particles.  The average $\Reff$
of the R99 data set is $\sim 13''$, corresponding to 3.34 kpc.  In the
merger remnants, 1 $\Reff$ is 3.5 kpc: this distance is chosen to be
$10''$ for the comparison with observations.  We use a slit width of
$2.5''$, \ie 0.875 kpc. After binning, we have 21 pixels on the
major axis and 16 on the minor axis. The bin size increases with
distance from the center, according to the observations of R99.

In each binned pixel, we construct the histograms of the
line--of--sight velocities ($\vlos$) of all the luminous particles
whose projected coordinates on the sky ($x^\prime$ and $y^\prime$)
fall within the pixel boundaries. This quantity is called the velocity
profile (VP). Subsequently, we parametrize the VPs using
Gauss--Hermite series (van der Marel \& Franx 1993, Gerhard 1993). We
checked that the kinematic parameters of each VP ($v, \sigma, h_3$ and
$h_4$) do not depend on the choice of the velocity bin size.
Furthermore, to increase the signal to noise and since most remnants
are close to an axisymmetric shape (in the equatorial plane, the
average axis ratio at 1$\Reff$ is roughly 0.8--0.9), we average the
results over 10 different position angles of the slit between
$0^{\deg}$ and $90^{\deg}$ in the equatorial
plane. Figure~\ref{f:fig_all_v_over_sigma_edge_on} and
\ref{f:fig_all_h3_edge_on} demonstrate that the results of this paper
still hold if we considered individual projected profiles, rather than
averaging 10 position angles together.

We estimate the errors on each kinematical quantities by
bootstrapping. For each spatial bin, we generate 100 bootstrapped VPs and
recompute the GH--decomposition. The error bar is then estimated as
the variance amongst the 100 bootstrapped results.

 In Figure~\ref{f:fig_v_sig_h3_h4_edge_on}, we show the basic
kinematic output of our simulations, the first four moments of the
VPs, \ie $v, \sigma, h_3, h_4$ and the degree of ordered motion
$v/\sigma$ on the major axis for the high resolution simulation
(266666 luminous and dark particles) of case 1 (see
table~\ref{t:initial_angles}).  The initial inclination angles of the
two disk galaxies are $[i_1,\omega_1,i_2,\omega_2] =
[-30,-30,30,30]$. The curves have been folded onto the positive side
of the major axis (with a sign change for the odd moments $v$ and
$h_3$) . In this way, we can estimate the left/right asymmetries.  In
most geometries analysed in this paper, the left and right side of the
major axis are similar within the errors. For this model, the mean
velocity increases linearly until 1 $\Reff$ and then rises more slowly
to reach 150 $\kms$ at 3 $\Reff$. In other geometries, the rise is
more monotonic but reaches the same value at 3 $\Reff$. The velocity
dispersion decreases from central values of 140--160 $\kms$ to 80--100
$\kms$ at 3 $\Reff$. The innermost point shows a lower value (by $\sim
20 \kms$) compared to its immediate neighbors. BB
observed the same behavior in their simulations and interpreted it as
follows: all the merger remnants have a large fractions of particles
from the Hernquist bulges of their progenitors in their inner
regions. These particles still follow a $r^{-1}$ density profile at
small radii and therefore have velocity dispersion that scale as
$r^{-1/2}$, producing the central dip.

 In Figure~\ref{f:high_low_resolution_v_over_sig} and
 \ref{f:high_low_resolution_h3}, we show the $v/\sigma$ and $h_3$
 profiles of two cases, where each simulation was rerun with three
 times more particles.  The low resolution simulations (with 88887
 particles) do not show significantly different results compared to
 the high resolution cases. In case 1, the low resolution models have
 more left/right asymmetry, but if one takes the average between the
 two sides, then the $v/\sigma$ and $h_3$ profiles are very similar.
 Therefore in the remainder of this paper, we will explore various
 initial conditions using only low resolution simulations.
 Similarly, Figure~\ref{f:high_low_resolution_h3} displays the $h_3$--profiles
 for the high/low resolution simulations: $h_3$ is typically 
 zero or positive inside 1 $\Reff$ (see also BB). 

\section{Discussion}
\label{s:discussion}

\subsection{Edge-on remnants}

 Figure~\ref{f:fig_edge_on_with_Bendo} overplots the observed
 $v/\sigma$ profiles of the R99 sample and the $v/\sigma$ profiles of
 all our merger models (see table~\ref{t:initial_angles}) when viewed
 edge--on.  In the models, the left and the right sides have been
 averaged.  The $v/\sigma$ profiles of the remnants can reach values
 between 1 and 2.2 in the very outer parts \ie at 3 $\Reff$. At
 smaller radii, \eg at 1 $\Reff$, they attain only [0.1, 1.0],
 depending on the geometry. BB find similar results for their 3:1
 merger remnants, as indicated by the shaded region in Figure
 ~\ref{f:fig_edge_on_with_Bendo}. The R99 data set shows higher values
 of $v/\sigma$ at all radii: at 1 $\Reff$ $v/\sigma_{\rm observed}$
 covers the range [0.8, 2.0] and [1.4, 3.5] at 2 $\Reff$. Note also
 the sharp increase in $v/\sigma_{\rm observed}$ at small radii (R
 $\le 0.5 \; \Reff$). Furthermore the merger simulations do not
 reproduce the central peak in $\sigma$ observed by R99 (see their
 Figure 1). From this comparison alone we could conclude that
 dissipationless simulations can not reproduce the observed kinematics
 of the R99 sample. In the next section, we will also use the
 GH--moment $h_3$ and the minor axis rotation to compare with the
 observations.

\subsection{Inclined remnants}

 When viewed edge-on, our merger remnants have an apparent ellipticity
 $\epsilon_{\rm remnant} \simeq 0.5 - 0.6$ at 1 $\Reff$, much more
 flattened than the objects of R99 with $\langle \epsilon_{\rm
 observed} \rangle = 0.3$ (see their Figure 3). In order to make a
 fair comparison to this dataset, we need to incline our merger
 remnants such that they have the same ellipticity $\epsilon_{\rm
 remnant} \simeq 0.3$ at 1 $\Reff$. As can be expected, the mean
 line--of--sight velocity is lowered (compared to the edge--on
 case). The velocity dispersion is also diminished, but not as much as
 the velocity. Therefore the $v/\sigma$ profiles of the inclined cases
 do not rise as fast as the edge--on cases
 (Figure~\ref{f:fig_obs_and_models}).

The $h_3$--profiles are less sensitive to inclination, \ie they are
similar to those of the edge-on case. In
Figure~\ref{f:h3_obs_and_model}, we only show the $h_3$ values of the
merger remnants in the inclined case.  These profiles are essentially
zero (or slightly positive) inside 1 $\Reff$, whereas the observed
$h_3$ values show an outward decline towards $\langle h_3 \rangle
=-0.06$ within the same radial interval. The model values reach 
--0.1 only at very large radii (3 $\Reff$).  Bender, Saglia \& Gerhard
(1994) observed a sample of 44 elliptical galaxies and found $\langle
h_3 \rangle = -0.1$ for the disky ($a_4/a = 0.02$) objects (see their
Figure 14a). We choose the disky subsample, since they correlate with
the low--luminosity ellipticals. Their low values of $h_3$ is roughly
consistent with the R99 profiles, since Bender, Saglia \& Gerhard
obtained spectra only inside $1/2 \; \Reff$. 

\subsection{Minor axis rotation}

We have computed the normalized minor axis rotation, defined as
$v_{\rm min. norm.} =$\\$v_{\rm minor} / \sqrt{v^2_{\rm major} +
\sigma^2_{\rm major}}$ for all merger models and observations.  Note
that the normalization by the total kinetic energy avoids a division
by zero in the center. We find that both the merger remnants and the
galaxy sample observed by R99 have only a small amount of minor axis
rotation: $\langle v_{\rm min. norm.}\rangle_{\rm merger} = 0.04$ and
$\langle v_{\rm min. norm.}\rangle_{\rm R99} = 0.08$. The spatial average
has been done inside 2 $\Reff$. 

\subsection{Comparison with S0 galaxies}

 Fisher (1997) observed a sample of 18 S0 galaxies and derived
 GH--moments along their major and minor axis. In most cases, these major
 axis $h_3$ profiles show a similar behavior than observed in our
 merger remnants: going from the center to the outer parts, $h_3$
 values increase from zero to some positive values, then decrease and
 change sign to finally remain negative. The maximum value of $h_3$
 reached by the data is in the range 0.05--0.1. However, the change of
 sign occurs at (or within) 1 $\Reff$ for these S0s, whereas it is
 between 1 and 2 $\Reff$ for the merger remnants. Furthermore, the
 shape of the S0s velocity curves is very different compared to the one of
 the merger remnants (see Figure 10 of Fisher) and their $v/\sigma$
 values are even higher than the low--luminosity ellipticals.

\subsection{Kinematic temperature of the initial galaxy disk}

 The merging process heats up the disks (\ie increases $\sigma$), 
decreases $v$ and therefore lowers the $v/\sigma$ profile. Among other
comparisons in this paper, we reject the merging scenario because we
were unable to reproduce the observed $v/\sigma$ values, but this
scenario might be viable if the initial disks were colder. Therefore
it is interesting to know the {\it initial} temperature of the disk
before the merger. In Figure~\ref{f:initial_disk_v_over_sig} (bottom
panel), we show this initial $v/\sigma$ profile of the largest disk
galaxy. It rises linearly with radius and reaches $v/\sigma$ = 7.5 at
3 disk scale length along the major axis (edge--on view). The initial
disk galaxy has a bulge to disk mass ratio of 1/3, so it compares to a
Sb galaxy (see \eg Figure 4.51 of Binney \& Merrifield 1998). For
comparison, the old stellar population of Milky Way galaxy (SBb)
reaches $v/\sigma \simeq 4$ at the Sun position (2.5 disk scale
length). We computed the $v/\sigma$ profile for six Sa galaxies from
Corsini \etal (1999); they reach $v/\sigma = 2$ at one disk scale
length ${\rm R_{exp}}$ and $v/\sigma = 4$ at two ${\rm R_{exp}}$ with
large error bars in the outer parts. The inclination $i$ does not play
a role here since $v/\sigma$ is almost independent of $i$, when $i \le
50$ away from edge--on. Bottema (1999) has measured velocity and
velocity dispersion for the Sb galaxy NGC 7331 up to 3 $\Reff$. The
$v/\sigma$ profile of that galaxy reaches values around 8 at 3
$\Reff$, but has a steeper inner slope: at 1 $\Reff$, NGC 7331 has
$v/\sigma=4$ already. We have started a merger simulation with a
colder initial disk using the geometry of case 1 (see
table~\ref{t:initial_angles}).  We find virtually the same $v/\sigma$
profile in the remnant of the merger simulation (see
Figure~\ref{f:initial_disk_v_over_sig}, top panel).  We conclude that
our choice of initial galaxy disk is not an artificial cause for the
low $v/\sigma$ profile found in our merger remnants.

\section{Could small ellipticals still be merger products?}
\label{s:remedies}

In this section, we speculate on some possible scenarii that could
still save the 3:1 merger of disks as a viable route for the formation
of fast rotating ellipticals. In addition, one (non--merger) mechanism is
mentioned that could also lead to the formation of similar objects.

If we start with a sub--maximal initial disks, a weaker bar
instability occurs since massive dark halos tend to stabilize disks
against bars. Therefore the disks are less heated during the merger
and could produce a remnant with a higher $v/\sigma$ profile, in
agreement with the R99 sample. Using a fast--rotating bulge in the
initial spiral galaxy could increase the $v/\sigma$ profile of the
remnant in the inner parts. Finally, the inclusion of a dissipative
component in the merger is also likely to increase the amount of
ordered motion: in gas--rich 3:1 merger simulations, large ($\sim 1
\;\Reff$) gaseous disks form in the remnant (Naab 2000). If new stars
are subsequently formed in such disks, the final rotational support is
increased. It is however not clear what fraction of the gas is turned
into stars (and under what conditions and timescale).

In a cluster environment, mergers occur rarely since the velocity 
dispersion is too high. However, galaxy harassment is likely 
to play a significant role. According to Moore, Lake \& Katz (1998)
disks galaxies in rich clusters undergo a complete 
morphological transformation from "disks" to "spheroidals". These 
objects probably retain a large amount of rotation and may therefore
have a high $v/\sigma$. 

Finally, we have compared our merger remnants with the low--mass end
of the elliptical sequence. Intermediate ellipticals ($-19 < {\rm M_B}
< -20$) show more modest $v/\sigma$ profiles.  The kinematics of 
these objects may still be explained by 3:1 collisionless mergers.

\section{Conclusions}

We have compared various kinematic characteristics of a limited sample
of 3:1 N--body merger remnants with observations of low--luminosity
elliptical galaxies in order to test if the collisionless merger of
disk galaxies (with mass ratio of 3:1) are a likely formation
mechanism for such objects. The simulated merger remnants and the
observed galaxies both show rapid rotation along the major axis and
little rotation along the minor axis. A detailed comparison between
simulations and data shows, however, that on average, the rotational
support $v/\sigma$ of the simulated remnants is too small compared to
the observed galaxies. A comparison of the line--of--sight VP,
quantified by $h_3$, reveals that the merger remnants have VPs with
zero or positive $h_3$ inside 1 $\Reff$, whereas the observed VPs have
$\langle h_3 \rangle \simeq -0.06$ in the same interval. These more
direct comparisons with the data support the conclusion of R99 based
on simpler models, \ie the low--luminosity galaxies have a different
dynamics than the 3:1 merger remnants. In particular they have, on
average, more ordered motion.  These conclusions differ from the ones
reached by BB, even though the results of our simulations agree with
theirs. The difference can be traced to the more detailed and rigorous
data model comparison performed here.

At face value this suggests that either the last merger that this
class of small "ellipticals" experienced had a mass ratio $>$ 3:1, or
that a substantial stellar disk formed by gas dissipation after the
last major merger. Note that we have been conservative by looking at
3:1 mergers in the sense that the 3:1 remnants are the most likely
objects to look like an elliptical and have a high rotation support:
1:1 mergers lead to boxy objects with far less rotation support and
5:1 mergers (or higher) do not form an elliptical because the large
disk mostly survive the interaction. It is not clear
quantitatively what is the effect of the inclusion of gas and star
formation in the kinematics of the remnant. Furthermore, adding gas
might not be the only way to save the merging scenario. For example,
starting the merger with a sub--maximal disk could produce a remnant
with a higher $v/\sigma$ profile. On the other hand, galaxy harassment
is a plausible mechanism to heat the disk of spirals, while keeping
some degree of ordered motion. However harassment is mainly operating
in rich clusters, whereas mergers are ubiquitous in a hierarchical
formation scenario.

%------------------------------------------------------------------------
% Acknowledgments 

\acknowledgments
  
We thank Josh Barnes for his constructive comments on the manuscript. 

%
% References
%

{}

\clearpage

\begin{deluxetable}{rrrrr}
\tablecolumns{5} 
\tablewidth{0pc}
\tablecaption{Angles specifying the initial conditions for the 
two disk galaxies \label{t:initial_angles}}
\tablehead{
\colhead{$i$} &
\colhead{$i_1$} &
\colhead{$\omega_1$} &
\colhead{$i_2$} &
\colhead{$\omega_2$} \\
\colhead{(1)} &
\colhead{(2)} &
\colhead{(3)} &
\colhead{(4)} &
\colhead{(5)} \\
}
\startdata
1 &  --30 &  --30  &   30 &    30 \\
2 &   90 &    0  &   0  &    0  \\
3 &    0  &   0  &   0  &    0  \\
4 &  180  &   0  &   0  &    0  \\
5 &  --71  &  90  & 109  &   90 \\
6 &   30  &   0  &   0  &    0 \\
7 &   60  &   0  &   0  &    0 \\
8 &  --90  &  90  &  90  &   90 \\
9 &  --30  &   0  &   0   &   0  \\
10 &  --60  &   0  &   0  &    0 \\
11 &   60  &  60  & --60  &  --60 \\
12 &  120  &   0  &   0  &    0 \\
13 &  150 &    0  &   0  &    0 \\
14 & --120 &  --30  &  60  &  --30 \\
15 &   60 &   --30 & --120 &   --30 \\
16 &   60 &   30  & --60  &  --30 \\
\enddata
\tablecomments{Column~(1) gives the number of the simulation, 
  column~(2) the angle between the spin plane and the 
orbital plane for the massive disk galaxy and column~(3) the angle 
between the line of nodes and the pericenter. Column~(4) and (5) 
are the same than (2) and (3), but for the small spiral galaxy.}
\end{deluxetable} 

\clearpage 

%------------------------------------------------------------------------
% Print Figures 

\plotone{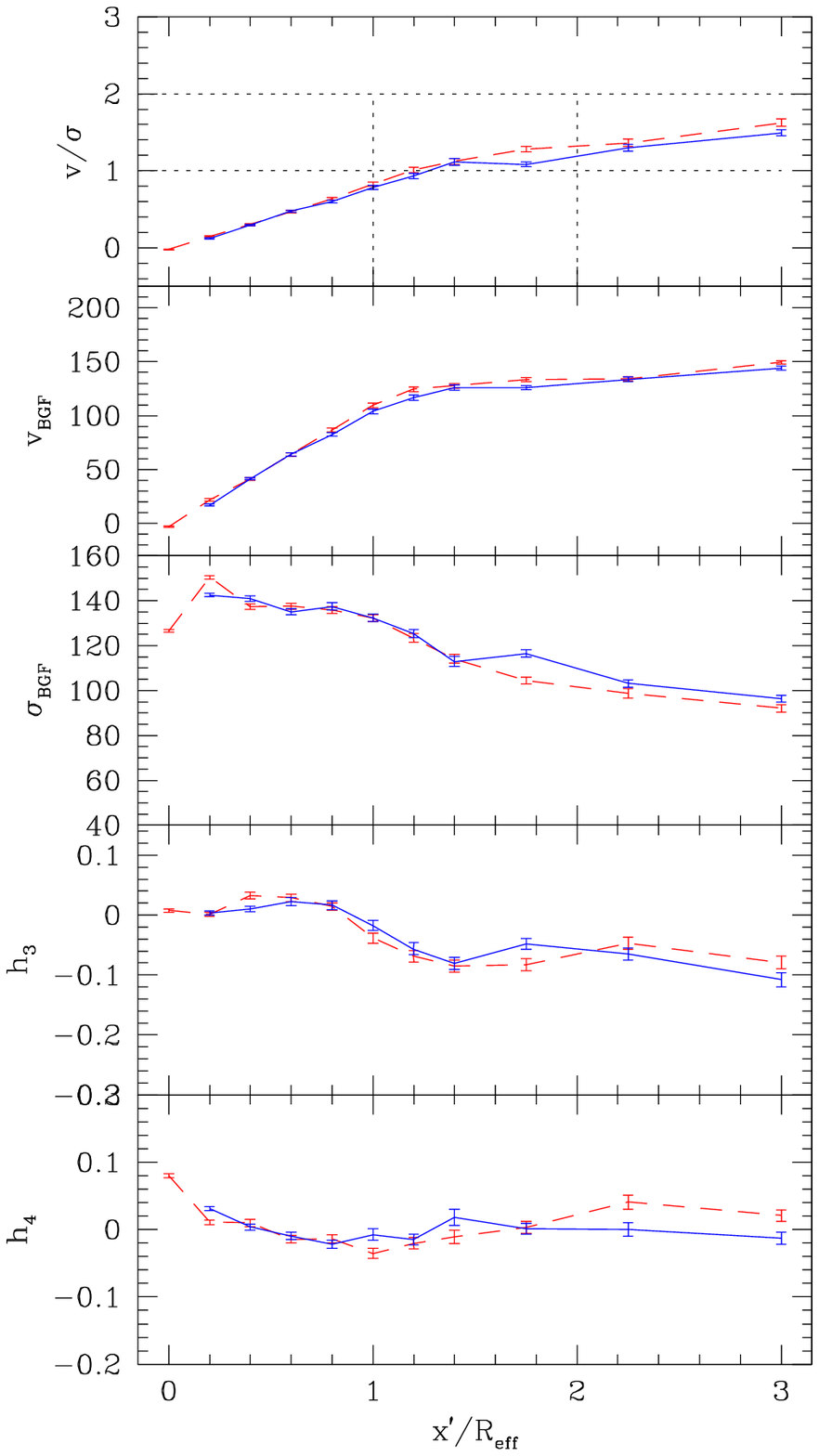}
\figcaption[fig_v_sig_h3_h4_edge_on]{First four moments of the GH
decomposition along the major axis of one merger remnant (case 1, see
table~\ref{t:initial_angles}): mean line--of--sight velocity $v$ (in km/s),
line--of--sight velocity dispersion $\sigma$ (in km/s) and GH--moments $h_3$ and
$h_4$. The top panel shows the rotational support, \ie the ratio
$v/\sigma$. The remnant is viewed edge--on here. The left side of the remnant 
(dashed line) has been folded onto the right side to estimate the asymmetry. 
\label{f:fig_v_sig_h3_h4_edge_on}}

\plotone{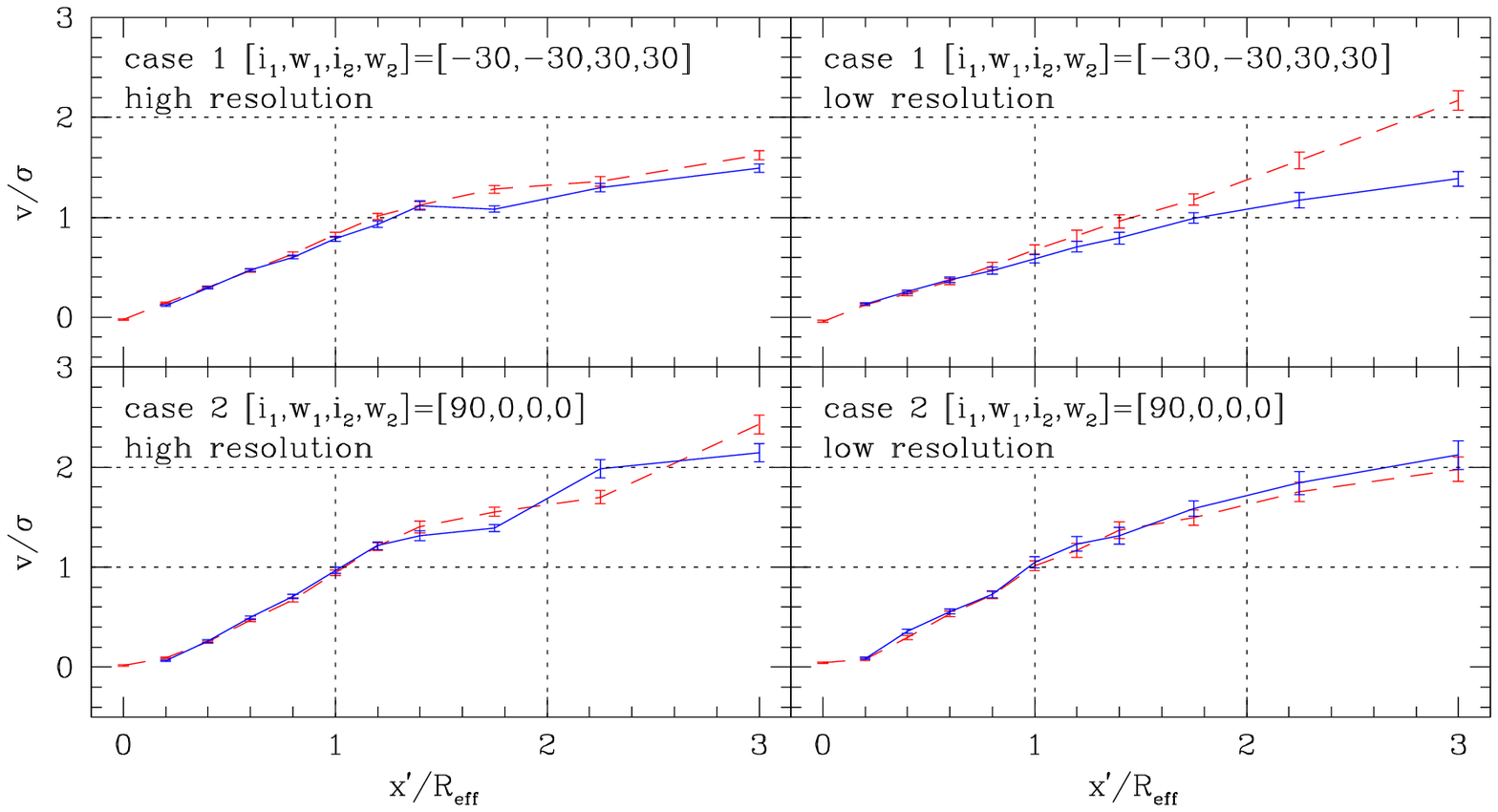}
\figcaption[high_low_resolution_v_over_sig]{$v/\sigma$ profiles for
the first two geometries.  The left panels show simulations with three
times more particles than those in the right panels. All profiles are
viewed edge--on. \label{f:high_low_resolution_v_over_sig}}

\plotone{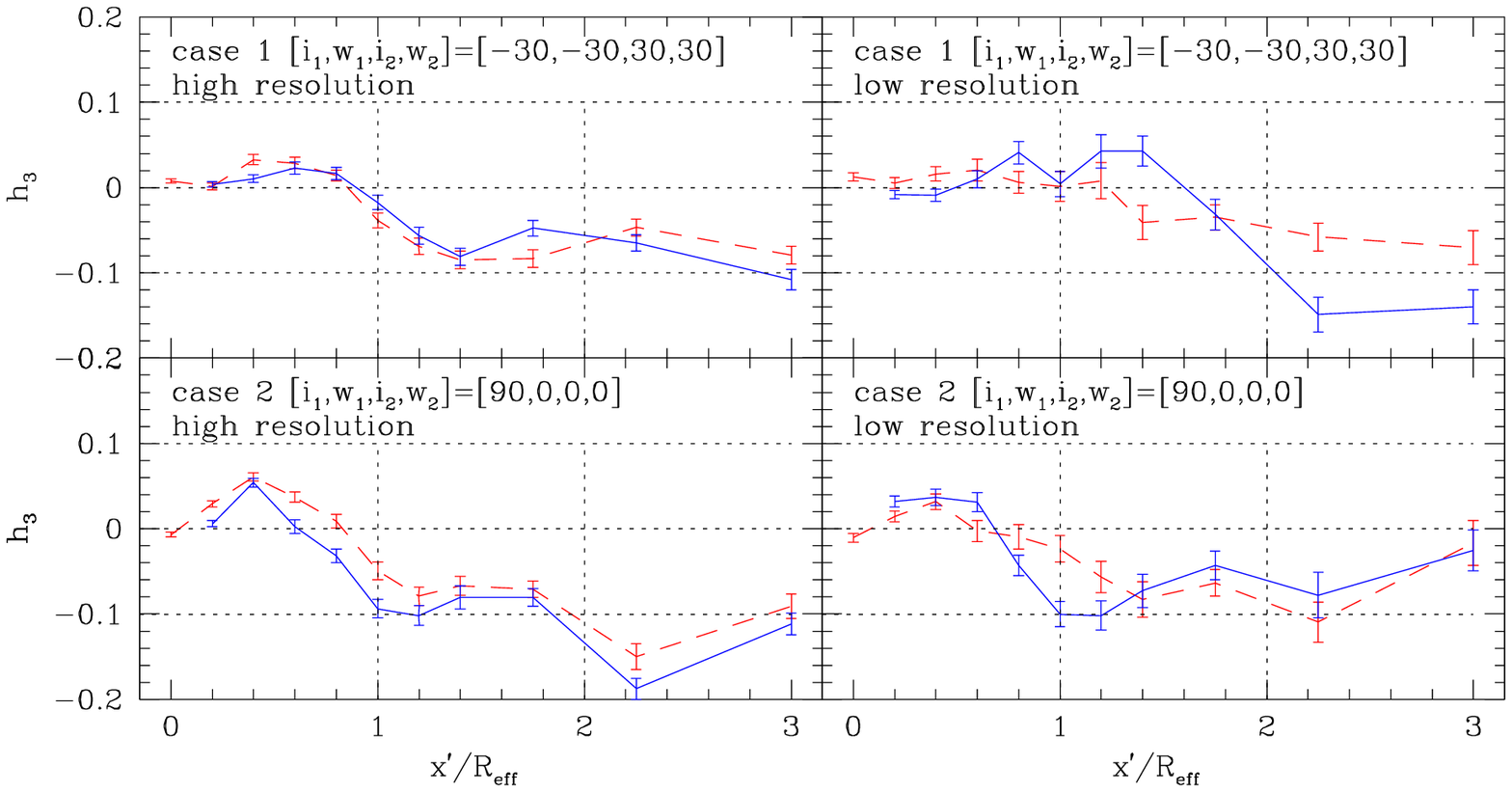}
\figcaption[high_low_resolution_h3]{Same as
Figure~\ref{f:high_low_resolution_v_over_sig}, but for
$h_3$. \label{f:high_low_resolution_h3}}

\plotone{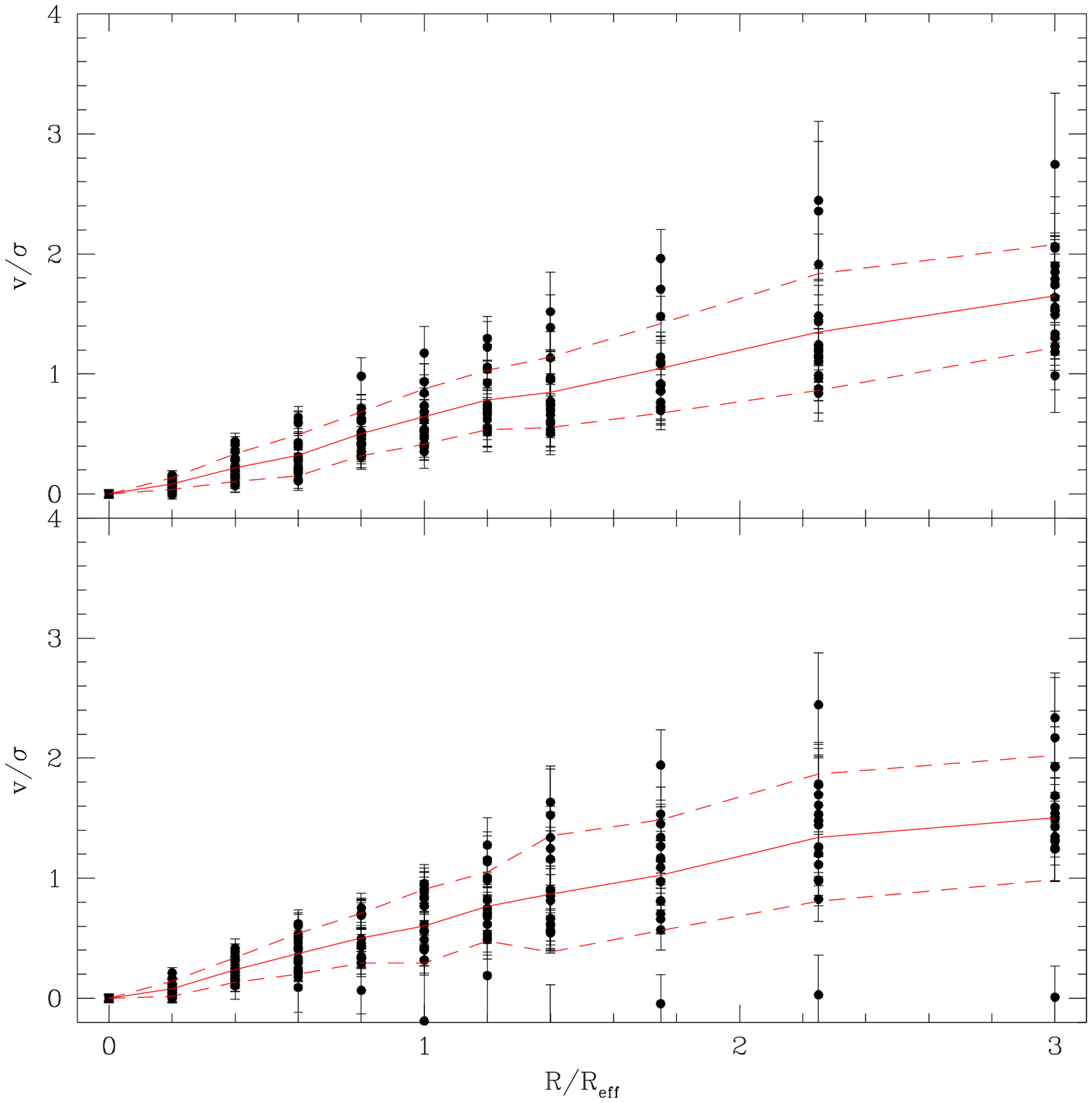}
\figcaption[fig_all_v_over_sigma_edge_on]{
$v/\sigma$ profiles viewed edge--on for all the merger 
simulations. Here no azimuthal average is performed: the bottom panel
shows the results measured along the $x$--axis and the top panel along
the $y$--axis in the equatorial plane. The full line is the average 
profile, while the two dashed lines show the $1\sigma$ band. On average
both distributions are identical, so we conclude that 
the results are independent of the viewing angle.
\label{f:fig_all_v_over_sigma_edge_on}}

\plotone{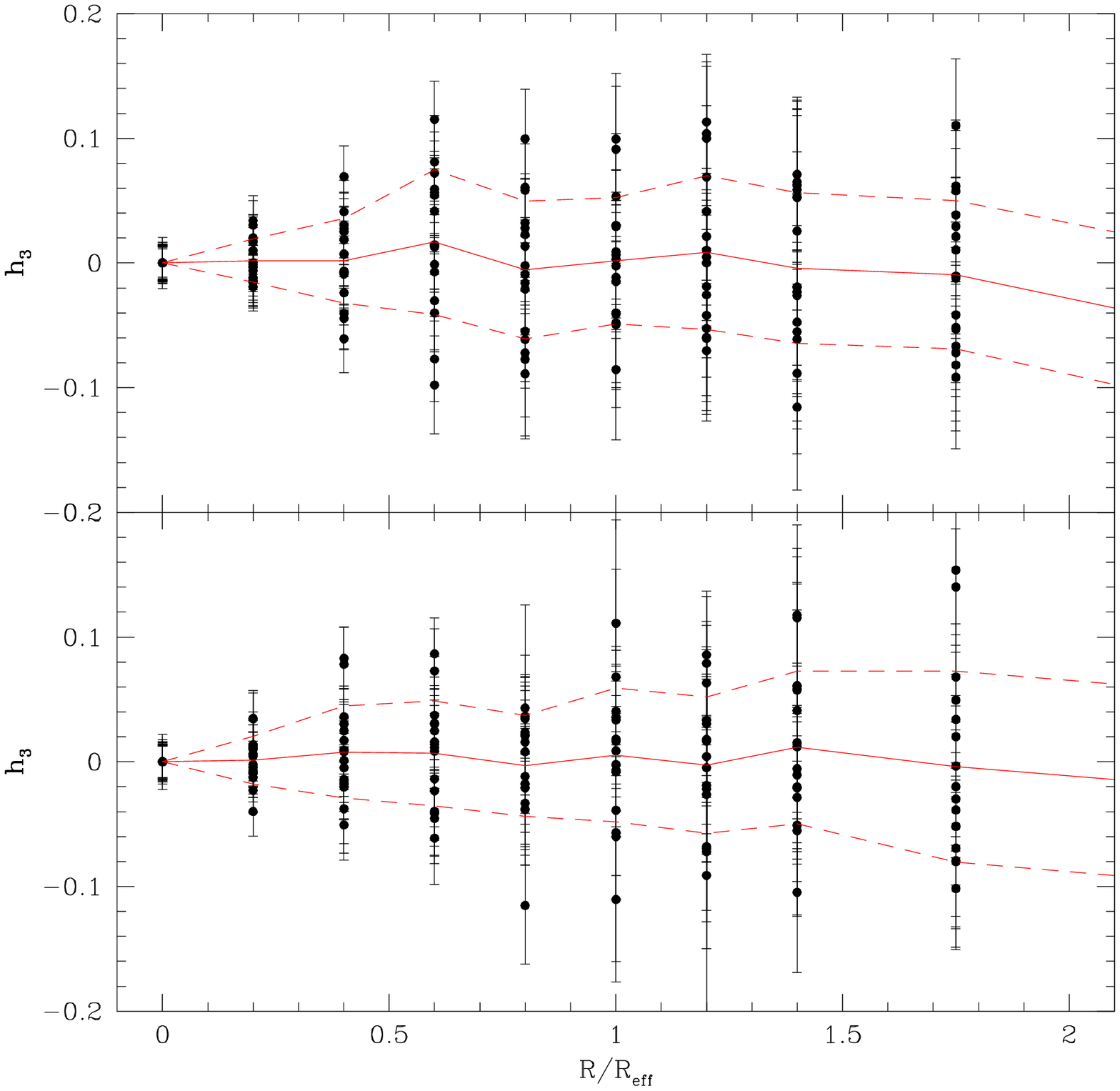}
\figcaption[fig_all_h3_edge_on]{Same as
Figure~\ref{f:fig_all_v_over_sigma_edge_on}, but for
$h_3$. \label{f:fig_all_h3_edge_on}}

\plotone{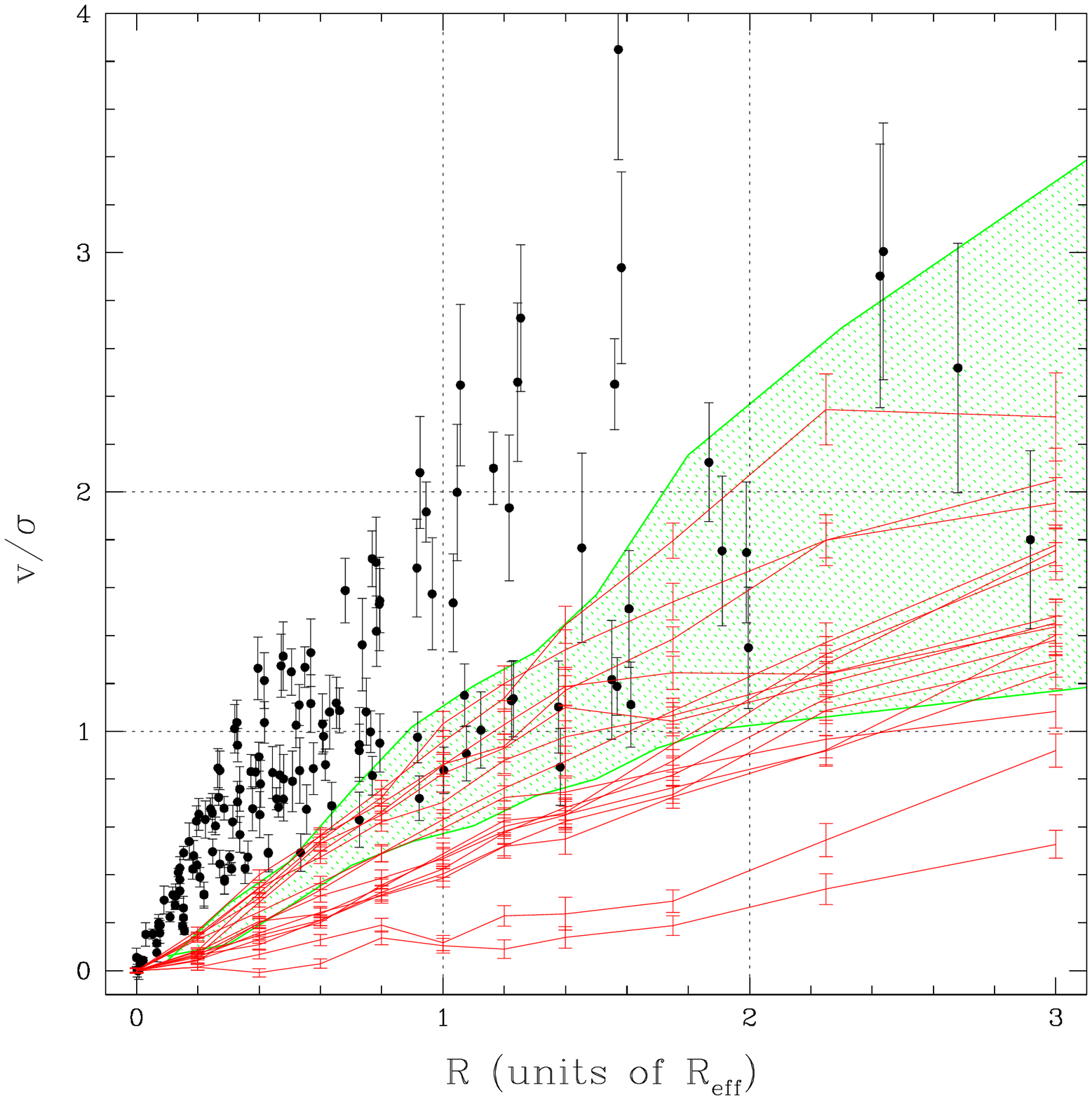}
\figcaption[fig_edge_on_with_Bendo]{Comparison of observed $v/\sigma$
(filled dots with error bars) and $v/\sigma$ for the edge--on merger
remnants (lines with error bars). The shaded area corresponds to the
range occupied by the edge--on models of BB. Both
distributions are not compatible with the observed one, even in this
edge--on case.  \label{f:fig_edge_on_with_Bendo}}

\plotone{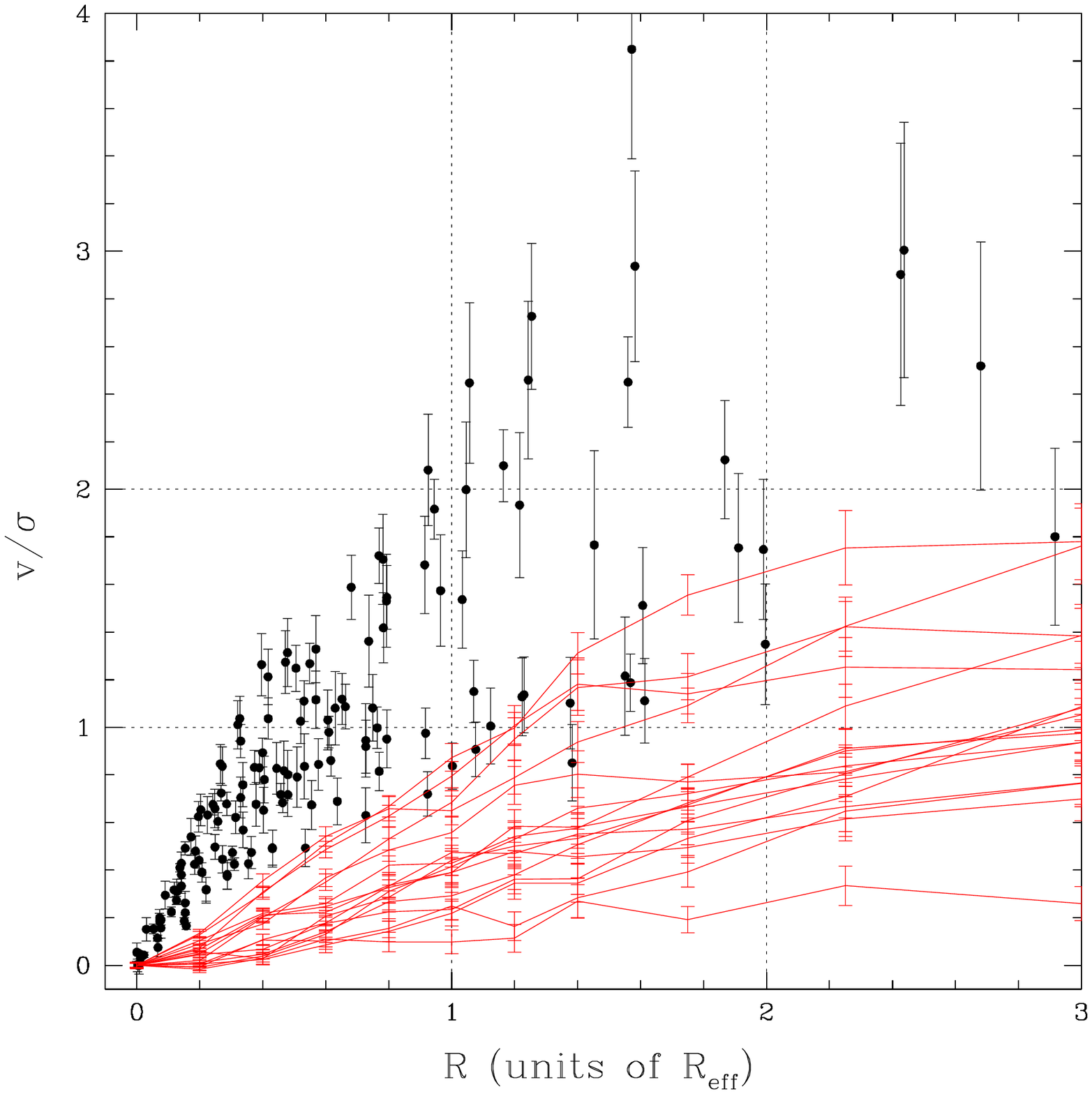}
\figcaption[fig_obs_and_models]{Same as
Figure~\ref{f:fig_edge_on_with_Bendo}, but with inclined merger
remnants. The inclinations are chosen such that each merger remnant
has an apparent $\epsilon=0.3$, which is the average value of the R99
sample.
\label{f:fig_obs_and_models}}

\plotone{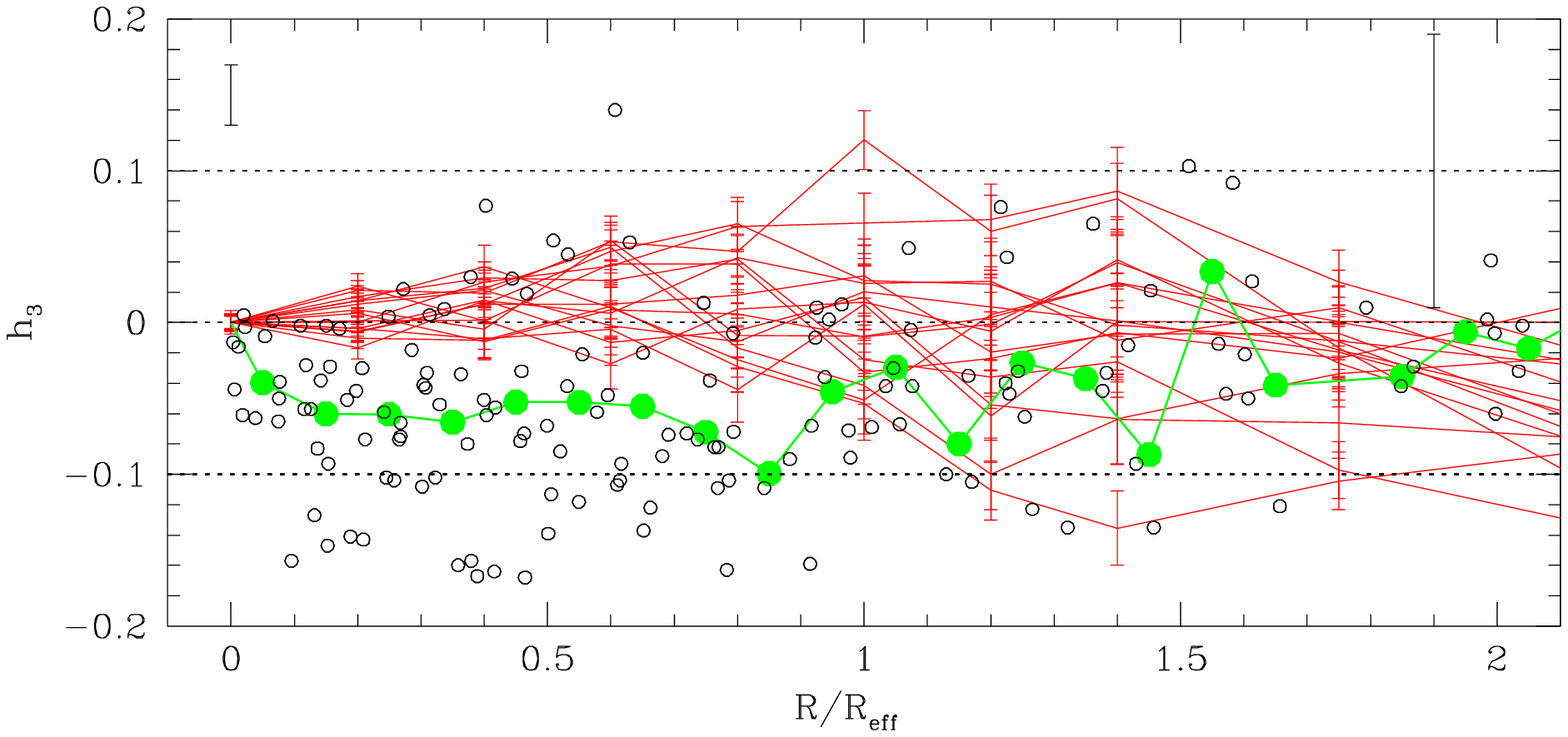} 
\figcaption[h3_obs_and_model]{
Same as Figure~\ref{f:fig_obs_and_models}, but for $h_3$. The big filled dots
are the average profile of the observations and the small open dots are 
the observed $h_3$ data points. Outside 1 $\Reff$, the error bars
become too large to make a meaningful comparison. A typical error bar is 
reproduced in the upper left (for the center) and right (for the outer parts)
corners. 
\label{f:h3_obs_and_model}}

\plotone{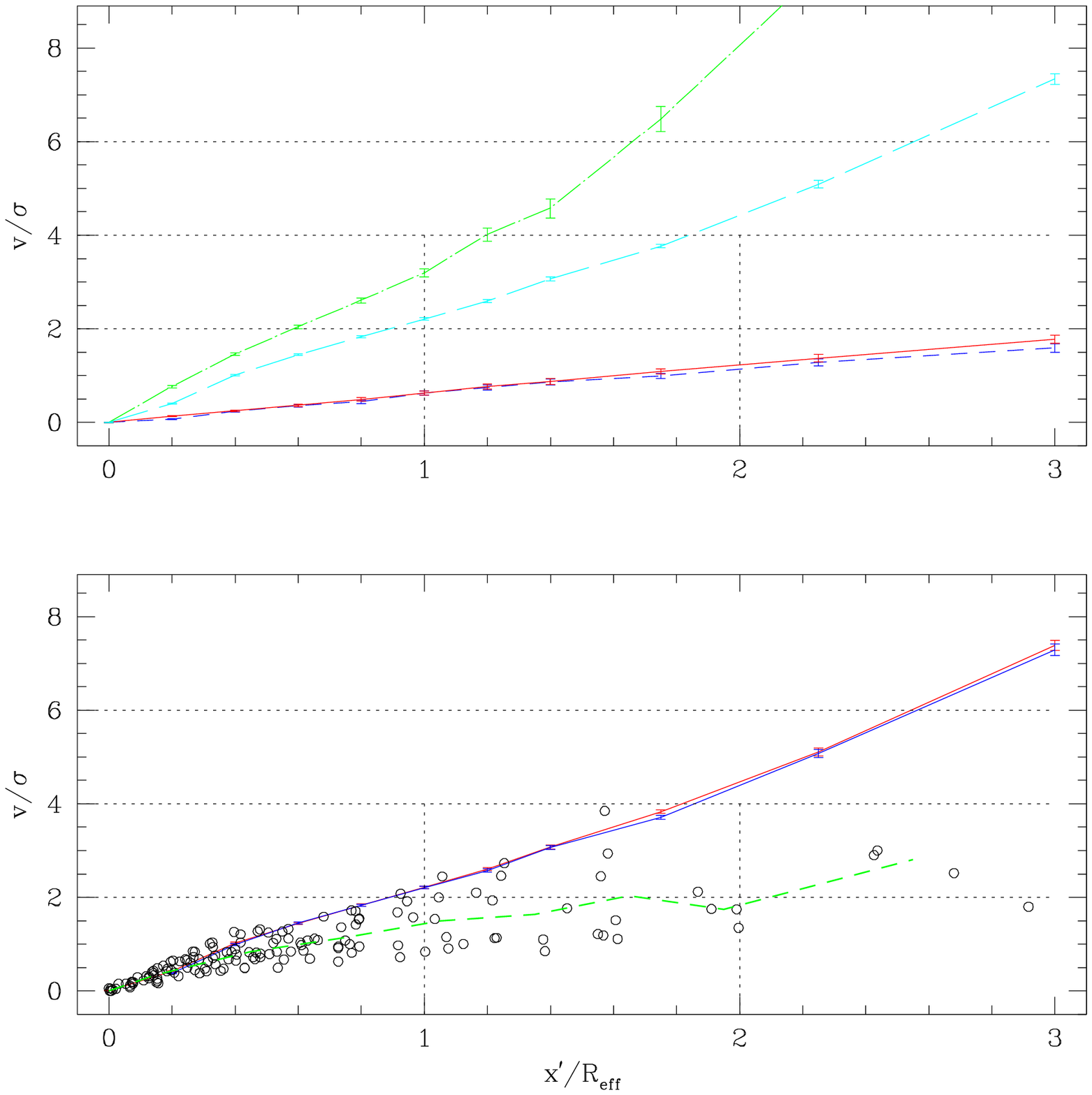}
\figcaption[initial_disk_v_over_sig]{Bottom panel: temperature
$v/\sigma$ of the initial disk galaxy viewed edge--on (full line). The
open dots are the individual $v/\sigma$ of the R99 sample, while the
dashed line is their mean $\langle v/\sigma \rangle$ profile. Top
panel: temperature $v/\sigma$ of a cold disk (long dash--dotted line)
and of our standard disk (long dashed line, see also bottom
panel). The short--dashed line is the $v/\sigma$ profile of the
merger remnant initiated with this cold disk.  For comparison, the
full line is the remnant $v/\sigma$ profile in case 1 (see
table~\ref{t:initial_angles}), which started from the (warmer)
standard disk.
\label{f:initial_disk_v_over_sig}}

\end{document}